%% file: BCl3_Annealing_JCP.tex
\documentclass[aip,graphicx]{revtex4-2}

\usepackage{chemformula} % Formula subscripts using \ch{}
\usepackage[T1]{fontenc} % Use modern font encodings
\usepackage{color}
\usepackage{amssymb,amsmath}
\usepackage{dcolumn}
\usepackage{multirow}
\usepackage{hyperref}
\usepackage{siunitx}
\usepackage{booktabs}
\usepackage{amssymb}
\usepackage{graphicx}
\usepackage{wasysym}
\usepackage{listings}

\begin{document}

\title{Reaction pathways of BCl$_3$ for acceptor delta-doping of silicon}
%\title{Optimal pathways for boron $\delta$-doping of silicon with BCl$_3$}

\author{Quinn Campbell}
\email{qcampbe@sandia.gov}
\thanks{Q.C. and K.D. contributed equally to this work}
\affiliation{Center for Computing Research, Sandia National Laboratories, Albuquerque NM 87185}

\author{Kevin J. Dwyer}
\thanks{Q.C. and K.D. contributed equally to this work}
\affiliation{Department of Physics, University of Maryland, College Park, MD 20742}

\author{Sungha Baek}
\affiliation{Department of Physics, University of Maryland, College Park, MD 20742}

\author{Andrew D. Baczewski}
\affiliation{Center for Computing Research, Sandia National Laboratories, Albuquerque NM 87185}

\author{Robert E. Butera}
\email{rbutera@lps.umd.edu}
\affiliation{Laboratory for Physical Sciences, College Park, MD 20740}

\author{Shashank Misra}
\affiliation{Sandia National Laboratories, Albuquerque NM 87185}

\date{\today}

\begin{abstract}
%Optimizing the electrical properties of $\delta$-doped layers formed on Si(100) requires a detailed understanding of adsorption and dissociation reaction pathways of dopant precursor molecules such as BCl$_3$, a promising candidate for atomic precision acceptor doping.
BCl$_3$ is a promising candidate for atomic-precision acceptor doping in Si, but optimizing the electrical properties of structures created with this technique requires a detailed understanding of adsorption and dissociation pathways for this precursor.
Here, we use density functional theory and scanning tunneling microscopy (STM) to identify and explore these pathways for BCl$_3$ on Si(100) at different annealing temperatures. 
We demonstrate that BCl$_3$ adsorbs selectively without a reaction barrier, and subsequently dissociates relatively easily with reaction barriers $\approx$1~eV.
Using this dissociation pathway, we parameterize a Kinetic Monte Carlo model to predict B incorporation rates as a function of dosing conditions.
STM is used to image BCl$_{3}$ adsorbates, identifying several surface configurations and tracking the change in their distribution as a function of the annealing temperature, matching predictions of the kinetic model well.
%We also predict a surface migration pathway active at annealing temperatures that leads to the formation of a stable, electrically inactive B-dimer complex. Electrical measurements of BCl$_3$-dosed $\delta$-layers show good agreement with the model, with carrier concentrations dropping upon annealing above \SI{250}{\celsius}. However, comparisons of the carrier concentration with SIMS measurements indicate that dimer formation is not significant here. SIMS also shows that an incorporation anneal is necessary to retain maximal B.
This straightforward pathway for atomic-precision acceptor doping helps enable a wide range of applications including bipolar nanoelectronics, acceptor-based qubits, and superconducting Si.
\end{abstract}

\maketitle

\section{Introduction}

Atomic precision advanced manufacturing (APAM) techniques are essential for the realization of a wide variety of novel quantum and electronic devices in Si such as dopant-based qubits,\cite{ward2020atomic,bussmann2021atomic} analog quantum simulators,\cite{khajetoorians2019creating} and superconducting devices.\cite{shim2014bottom}
APAM utilizes a Si surface that has been passivated with a monatomic resist such as H or Cl.\cite{pavlova2018first,dwyer2019stm,pavlova2021reactivity}
A precision lithography tool in the form of a scanning tunneling microscope (STM) is then used to depassivate a region of interest\cite{lyding1994nanoscale,randall2009atomic} that selectively reacts with a gaseous molecular precursor in ultrahigh vacuum (UHV).
Under appropriate conditions, the precursor will fully dissociate on the surface and ultimately incorporate a dopant atom into the Si substrate.
The carrier density achieved with APAM can readily exceed \SI{1e14}{cm^{-2}}, creating well-defined, quasi-2D $\delta$-doped regions within Si.\cite{keizer2015suppressing}
To date, APAM process development has focused almost exclusively on the formation of P-based devices using phosphine (PH$_{3}$) as the dopant precursor.\cite{schofield2003placement,oberbeck2004measurement,ruess2007realization}
However, the pursuit of APAM-compatible acceptor doping has gained interest for creating bipolar electronic devices,\cite{vskerevn2020bipolar} enabling qubits with high intrinsic spin-orbit coupling,\cite{ruskov2013chip,salfi2016quantum1,salfi2016charge} and the possibility of creating superconducting regions within Si.\cite{bustarret2006superconductivity,shim2014bottom,duvauchelle2015silicon,bonnet2021strongly}

Recent work has explored a variety of potential APAM-compatible acceptor precursors such as diborane (B$_2$H$_6$),\cite{vskerevn2020bipolar} aluminum trichloride (AlCl$_3$),\cite{radue2021alcl3} boron trichloride (BCl$_3$),\cite{Dwyer_BCl3} and organics such as trimethyl and triethyl aluminum\cite{owen2021alkyls}. Additionally, several solution-based approaches have been explored.\cite{ho2008controlled,ye2015boosting,silva2020reaction,frederick2021ultradoping,frederick2021reaction,silva2021solution}
Of these, BCl$_3$ in UHV has shown the most promise for $\delta$-doping applications, achieving an active carrier concentration of \SI{1.9e14}{cm^{-2}} without the need of subsequent thermal processing.\cite{Dwyer_BCl3}
Moreover, BCl$_3$ has been shown to be compatible with both H and Cl resists, enabling the fabrication of B-doped APAM nanowires\cite{Dwyer_BCl3} and fostering compatibility with PH$_3$ to create bipolar devices.\cite{vskerevn2020bipolar}
Further optimization of the BCl$_3$ doping process is likely to be achieved by fine-tuning the dosing and anneal parameters to maximize both dopant incorporation and activation. Consequently, a detailed understanding of the reaction pathways and kinetics can identify possible routes for improving incorporation results, and further understanding differences from the analogous process in PH$_3$ in creating devices.
This is particularly important in the context of achieving deterministic single-dopant incorporation.\cite{ivie2021impact}

In this work, we use Density Functional Theory (DFT) to elucidate the adsorption and dissociation pathway for BCl$_3$ on Si(100)--2$\times$1.
We demonstrate that BCl$_3$ adsorbs dissociatively onto the surface in a barrierless process, creating high levels of selectivity with both H and Cl resists.
Further dissociation occurs with barriers that are easily overcome near room temperature.
We feed these barriers into a Kinetic Monte Carlo (KMC) model to predict B incorporation levels as a function of dose temperature.
This contrasts with earlier work with phosphine which shows a several step dissociation pathway and higher reaction barriers for full dissociation.~\cite{Warschkow2016}
STM is also used to observe adsorbed BCl$_3$-related species (BCl$_x$ with $x$ = 1 to 3) on a Si(100) surface before and after annealing, with changes in surface structures found to be in good agreement with the dissociation behavior predicted by the model.
%As the surface is heated, the model predicts a surface migration pathway that leads to BCl fragments forming stable dimer complexes which reduce the electrically active carrier concentration.
% Electrical transport measurements of the active hole concentration in $\delta$-doped samples are found to agree qualitatively with the model, showing a gradual decrease in carrier concentrations for annealing temperatures above \SI{250}{\celsius}, predicted in the model to be due to BCl$_2$ desorption.
% The model further predicts a thermally-activated surface migration pathway resulting in BCl fragments forming stable dimer complexes that are electrically inactive.
% However, secondary ion mass spectrometry (SIMS) shows total B concentrations similar to hole concentrations, indicating B dimerization is not significant in these samples. B concentrations measured by SIMS are also below the total B levels predicted by the model suggesting some site blocking by other adsorbates from the vacuum.
% We also find that depositing Si before annealing surprisingly results in lower B levels, indicating an incorporation anneal is important for retaining B.
This work further confirms BCl$_3$ to be an ideal acceptor precursor for APAM processes with these refinements pointing the way toward room-temperature fabrication of bipolar nanoelectronic devices,\cite{vskerevn2020bipolar} acceptor based qubits,\cite{ruskov2013chip} and superconducting regions within Si.\cite{bustarret2006superconductivity}

\section{Methods}

\subsection{Electronic Structure Calculations}

We determine the thermodynamic adsorption energy of any particular BCl$_x$-surface configuration using DFT with the following equation:
\begin{equation}
	E_{\rm a} = E_{\rm slab/BCl_3} - E_{\rm slab} - E_{\rm BCl_3},
\end{equation}
where $E_{\rm a}$ is the adsorption energy of the molecule on the Si surface, $E_{\rm slab/BCl_3}$ is the total energy of the adsorbate on the slab, $E_{\rm slab}$ is the total energy of the slab without any adsorbate, and $E_{\rm BCl_3}$ is the total energy of the isolated BCl$_3$ molecule.
Negative values of $E_{\rm a}$ imply a thermodynamically favorable adsorption energy for that configuration.
All energy calculations are performed using the plane wave {\sc quantum-espresso} software package.\cite{Giannozzi2009}
To compute reaction barriers between configurations, we use the Nudged Elastic Band (NEB) method, also implemented in {\sc quantum-espresso}, and use norm-conserving pseudopotentials from the PseudoDojo repository\cite{VanSetten2018} and the Perdew-Burke-Ernzerhof exchange correlation functional.\cite{Perdew1996}
Kinetic energy cutoffs of 50~Ry and 200~Ry are used for the plane wave basis sets used to describe the Kohn-Sham orbitals and charge density, respectively, and a 2$\times$2$\times$1 Monkhorst-Pack grid is used to sample the Brillioun zone.\cite{monkhorst1976special}

Adsorption energy calculations are performed on the 4$\times$4 supercell of a seven-layer thick Si(100)--2$\times$1 slab with a 20~\AA~vacuum region, where a single unit cell has a size of 3.87~\AA $\times$ 3.87~\AA.
Matching the experimentally measured Si structure, we model the Si surface with alternating buckled Si dimers.
A H resist is placed on the surface with the exception of three Si dimer sites to gauge the selectivity of BCl$_3$ molecules on a bare Si surface versus a passivated surface.
On the other end of the slab, the Si dangling bonds are passivated with Se atoms to prevent spurious surface effects.
The bottom four layers of the slab are frozen and the geometry of the surface layers along with the adsorbate are relaxed until the interatomic forces are lower than 50~meV/\AA.
We compute the reference molecular energy for a single BCl$_3$ molecule in a 15~\AA$^3$ box.

\subsection{Kinetic Monte Carlo}

A KMC model\cite{Bortz1975,Gillespie1976} implemented using the KMCLib software package\cite{Leetmaa2014} is used to predict the incorporation rate of B atoms in patches of exposed Si(100) several nanometers wide.
Our KMC model uses transition rates based on the Arrhenius equation $\Gamma = A\exp{\Delta/k_{\rm B}T}$,\cite{Arrhenius1889} where $\Gamma$ is transition rate, $A$ is the attempt frequency, $\Delta$ is the reaction barrier found from our earlier DFT calculations, $k_{\rm B}$ is the Boltzmann constant, and $T$ is the temperature.
Attempt frequencies are set to $10^{12}$~s$^{-1}$ as a reasonable order of magnitude estimate based on an analysis of attempt frequencies for the dissociation of PH$_{3}$ on Si.\cite{Warschkow2016}

We calculate the effusive flow rate of molecules landing on any particular Si dimer as $\Phi_{effusion} = Pa/\sqrt{2\pi m k_{\rm B}T}$, where $P$ is the pressure of the incoming precursor gas, $a$ is the area of impingement, taken here as a single Si dimer, $m$ is the mass of the precursor gas, $k_{\rm B}$ is the Boltzmann constant, and $T$ is the temperature.

Each KMC calculation is repeated 200 times to obtain a meaningful statistical sampling of likely outcomes and we report the average outcomes, along with standard deviations as applicable. We calculate the standard error by assuming a Poisson distribution of measured counts.
For ease of reproducibility, we have placed our KMC code on github.\cite{kmccode}

\subsection{Si Surface Preparation}

BCl$_{3}$ dosing, annealing, STM imaging (ZyVector STM Lithography controller), and $\delta$-layer fabrication were performed in a ScientaOmicron VT-STM UHV system with base pressure $P$ $<$ \SI{2.7e-9}{Pa} (\SI{2.0e-11}{Torr}). Si(100) wafers used in these experiments were p-type (B-doped) obtained from ITME with resistivity $\rho$ = 1~$\Omega\cdot$cm to 10~$\Omega\cdot$cm. Samples 4~mm $\times$ 12~mm in size were cleaned by sonication in acetone, methanol, and isopropanol before being loaded into the UHV system on a ScientaOmicron XA sample plate. Clean Si(100)--(2$\times$1) surfaces were prepared in UHV by rapidly annealing the samples to \SI{1225}{\celsius} via Joule heating before being scanned in the STM for contaminants, as described elsewhere.\cite{Dwyer_BCl3}

\subsection{BCl$_{3}$ Dosing and $\delta$-Layer Fabrication}

Si(100) samples were dosed with BCl$_{3}$ within the UHV system immediately after initial annealing to form a clean surface. A precision leak valve was used to introduce BCl$_{3}$ into the UHV chamber and expose samples to a 1.5~langmuir dose (1~langmuir = 10$^{-6}$~Torr$\cdot$s) at room temperature ($\approx$\SI{20}{\celsius}) using a dosing pressure of \SI{4.0e-9}{Torr}. 
The samples were then imaged with STM both before and after anneals at different temperatures ranging from \SI{250}{\celsius} to \SI{600}{\celsius}.
% After dosing, two sets of samples referred to as ``ix'' and ``LLx'' were formed that received the same processing steps but in different orders. Those steps consisted of an anneal step between \SI{250}{\celsius} and \SI{1000}{\celsius} for 1~min (\SI{250}{\celsius} samples were annealed 30~min) and a short Si deposition step consisting of a 1~nm ``locking layer'' deposited without heating (described elsewhere\cite{keizer2015suppressing}). The ix samples received the anneal first to incorporate adsorbed B into the surface before the locking layer and are referred to specifically as i020 (no anneal), i250 (\SI{250}{\celsius}), i350 (\SI{350}{\celsius}), i400 (\SI{400}{\celsius}), i450 (\SI{450}{\celsius}), and i600 (\SI{600}{\celsius}). For each annealing temperature, two ix samples were dosed at the same time with one processed immediately and the other imaged with the STM before and after the incorporation anneal. The LLx samples received the Si locking layer deposition first before the anneal and are referred to specifically as LL250 (\SI{250}{\celsius}), LL350 (\SI{350}{\celsius}), LL400 (\SI{400}{\celsius}), LL450 (\SI{450}{\celsius}), and LL500 (\SI{500}{\celsius}). 
STM image processing and statistical analysis of adsorbates were performed using Gwyddion\cite{Gwyddion} to get feature coverages, reported in monolayers (ML), where 1~ML is equivalent to the Si(100) dangling bond density (\SI{6.78e14}{cm^{-2}}).

% To enable \textit{ex situ} analysis including SIMS and electrical measurements, BCl$_{3}$-dosed samples were immediately encapsulated with Si after the processing steps described above while remaining in UHV to form buried B $\delta$-layers. For encapsulation, $\approx$20~nm Si was deposited at a rate of $\approx$0.003~nm/s at $\approx$\SI{225}{\celsius} using an MBE Komponenten SUSI-63 sublimation source. Thermal radiation from the deposition source heated samples up to \SI{200}{\celsius} during this step. SIMS depth profiling was done externally by Eurofins EAG Materials Science.

\subsection{Electrical Characterization}

Electrical properties of B $\delta$-layer samples were determined using standard Hall bar magnetotransport measurements in a Quantum Design Physical Property Measurement System (PPMS). The Hall bars were patterned using electron beam lithography and were etched by a reactive ion etcher (RIE), as described elsewhere.\cite{Dwyer_BCl3} We created ohmic contacts using wire-bonding on mesa-defined pads. Each pad contained a 50 $\times$ 50 array of 1~$\mu$m sized etched squares and were wire-bonded with Al wires using a West Bond 7476E with a power of 300~W and a 30~ms time constant.

To study electrical carrier properties, samples were measured at 3~K while a magnetic field was applied perpendicular to the sample surface and swept from \mbox{--2.5~T} to +2.5~T. Four-terminal resistances were measured using an AC current source to apply 2~$\mu$A for longitudinal voltage measurements and 10~$\mu$A for transverse voltage measurements. 
%From these measurements, sheet resistance ($\rho_{\square}$), hole carrier concentration ($p$), and hole mobility ($\mu$) were extracted for the B $\delta$-layers.

\section{Results and Discussion}

\begin{figure}[t]
\includegraphics[width=\textwidth]{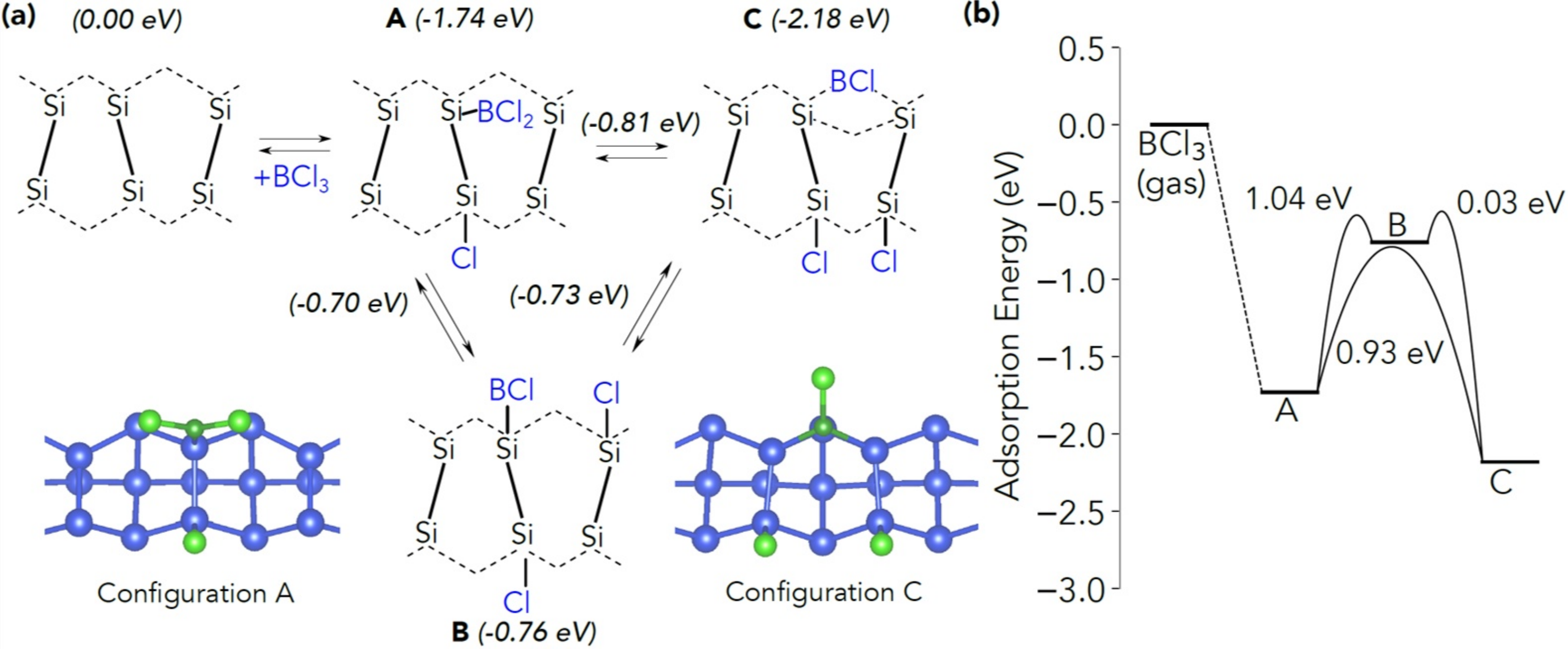}
\caption{(a) The reaction pathway of BCl$_3$ on a Si(100) surface. The numbers in parenthesis represent the adsorption energy of the configuration or the transition state (arrows).
The BCl$_3$ initially adsorbs without a barrier onto a single Si dimer, splitting so that BCl$_2$ takes up one side of the dimer and a Cl moves to the other side (configuration A).
The remaining Cl atoms are eventually shed until reaching configuration C where BCl bridges between two dimers, which we assume will lead to an eventual B incorporation. Configuration B is metastable. Si (blue), B (dark green), and Cl (light green) atomic positions are shown for configurations A and C.
(b) The energetic pathway for adsorption.
 \label{fig:rxn_pathway_schematic}}
\end{figure}

The thermodynamic reaction pathway for BCl$_3$ decomposing on a Si(100)--2$\times$1 surface was determined using DFT, with the lowest energy configurations shown in Fig.~\ref{fig:rxn_pathway_schematic}(a). A BCl$_3$ molecule adsorbs dissociatively onto the Si surface, with one Cl splitting off to one side of the Si dimer and the remaining BCl$_2$ bonding to the other side of the dimer (configuration A). Atomic positions for Si (blue), B (dark green), and Cl (light green) are shown for configuration A with the BCl$_2$ fragment residing over the dimer row. This initial reaction has an adsorption energy of \mbox{--1.74~eV}, indicating significantly stronger adsorption than is seen for PH$_{3}$ and B$_2$H$_6$, which have adsorption energies of around \mbox{--0.70~eV} to \mbox{--0.90~eV}.\cite{Warschkow2016,campbell2021model} This initial adsorption energy agrees well with recent theoretical work\cite{radue2021dopant} which estimates the BCl$_3$ adsorption energy on bare Si(100) to be \mbox{--1.76~eV} and matches previously predicted BCl$_3$ adsorption configurations.\cite{ferguson2009interaction}
Furthermore, the adsorption has no reaction barrier, implying a sticking coefficient of $\approx$1.
This correlates well with previous experiments that have demonstrated room temperature adsorption of BCl$_3$ on Si.\cite{ferguson2002atomic,consiglio2016comparison,pilli2018situ,Dwyer_BCl3}
This high adsorption energy can largely be attributed to the dissociation of BCl$_3$ and the formation of a strong Si-Cl bond: when BCl$_3$ bonds non-dissociatively on a bare Si surface, we calculate an adsorption energy of only \mbox{--0.28~eV}.
We also predict BCl$_3$ to be highly selective on both H and Cl resists, with an adsorption energy of \mbox{--0.03~eV} and \mbox{--0.02~eV}, respectively, corresponding to weak physisorption and matching previous reports of BCl$_3$ selectivity on these resists.\cite{Dwyer_BCl3}
This matches predictions that all chemical bonding configurations on the Si-Cl surface require overcoming significant reaction barriers of $\approx$2~eV.\cite{pavlova2021reactivity}

From this initial configuration, the BCl$_2$ fragment can proceed along one of two possible directions: losing a Cl to a nearby Si dimer (configuration B in Fig.~\ref{fig:rxn_pathway_schematic}), or moving directly to a BCl fragment bridging two neighboring Si dimers while a Cl moves to the opposite end of the neighboring dimer (configuration C).
Configuration B is metastable with an adsorption energy of \mbox{--0.76~eV}, and requires overcoming a 1.04~eV barrier from configuration A to move into, as shown in Fig.~\ref{fig:rxn_pathway_schematic}(b).
In contrast, configuration C is highly stable with an adsorption energy of \mbox{--2.18~eV}. Atomic positions for configuration C are also shown indicating the BCl fragment protrudes out between dimer rows.
Moving directly from configuration A to C has a barrier of 0.93~eV, implying that it will be more kinetically favorable for a BCl$_2$ fragment to move straight from configuration A to C instead of moving through the intermediate B state.
This reaction pathway significantly resembles the previously elucidated pathway for AlCl$_3$, although the reaction barriers are $\approx$0.3~eV higher.\cite{radue2021alcl3}

Overall, the reaction pathway is both straightforward in contrast with similar reaction pathways for PH$_{3}$ and B$_2$H$_6$,\cite{Warschkow2016,campbell2021model} and thermodynamically downhill. This pathway implies that if sufficient thermal energy is applied, a BCl$_3$ molecule can adsorb and move into a stable bridging configuration with no obstacles.

To predict the rate of B incorporation into a Si(100) surface as a function of the dose and annealing conditions (i.e. pressure, temperature, time), we use a KMC model based on our calculated reaction pathway.
We additionally include desorption reactions for configuration A, B, and C, although due to the high adsorption energy, these do not occur at low temperatures. 
Given the highly stable adsorption energy of the bridging BCl fragment (configuration C) and the significant reaction barrier to reversing the dissociation (1.37~eV), it is highly unlikely that it will be ejected from the surface in subsequent steps.
Within our kinetic model, we therefore consider reaching the bridging BCl fragment to be an incorporation event.
While this ignores subsequent steps of the incorporation process, previous kinetic models for PH$_{3}$ and B$_2$H$_6$ using analogous incorporation proxies of bridging PH and BH fragments have shown good agreement with experimental measures of incorporation.\cite{campbell2021model,ivie2021impact}
Based on the simplicity of the predicted reaction pathway and the low reaction barriers in the previous section, we expect that BCl$_3$ would lead to high levels of incorporation without the need for high temperatures.

\begin{figure}[t]
\includegraphics[width=0.5\textwidth]{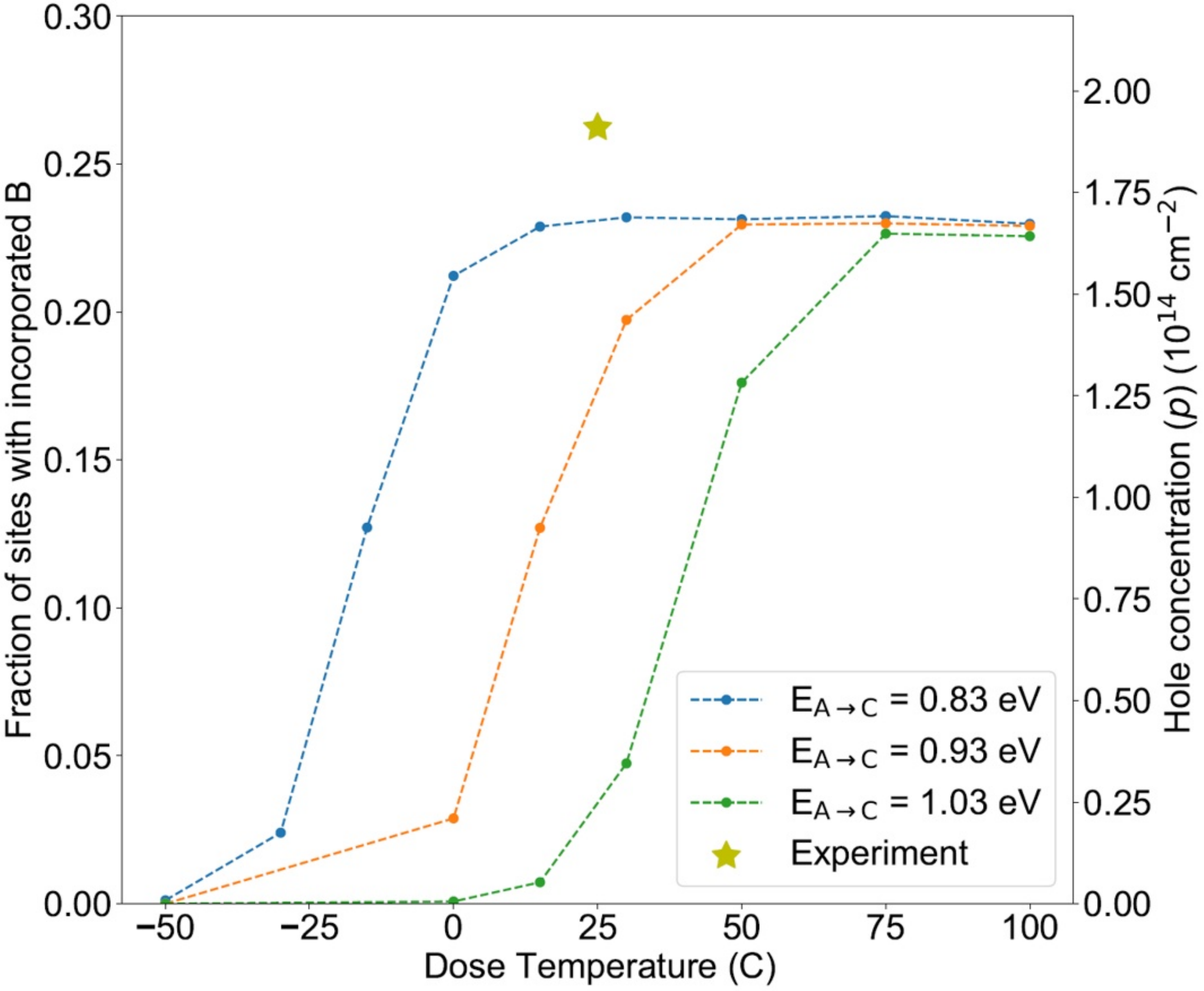}
\caption{KMC simulations of the incorporation fraction of BCl$_{3}$ on a 5~nm $\times$ 5~nm patch of bare Si(100) for various dose temperatures. We examine multiple values for the main barrier in the dissociation pathway (configuration A to C in Fig.~\ref{fig:rxn_pathway_schematic}). In all cases, saturation is achieved at relatively low temperatures ($<$\SI{100}{\celsius}). Also shown is an experimental value (star) of the measured hole concentration ($p$) for a $\delta$-layer sample dosed at room temperature (i020).
\label{fig:low-t-dosing}}
\end{figure}

Figure~\ref{fig:low-t-dosing} shows the kinetic model's predictions of the fraction of surface sites with incorporated B (bridging BCl) as a function of dose temperature.
This intuition is matched in our predicted results of incorporation in $\delta$-layer, with high incorporation rates occurring near room temperature with only minimal heating.
The onset temperature at which incorporation begins is highly dependent on the main reaction in the pathway connecting configuration A to C, as shown in Fig.~\ref{fig:rxn_pathway_schematic}.
Given that DFT reaction barriers are likely only accurate to $\pm$0.1~eV, we investigate the impact of raising or lowering the barrier by 0.1~eV.
For the previously calculated barrier (0.93~eV) the system reaches saturation of incorporated B (corresponding to roughly 0.25~ML of coverage) at a little over room temperature ($\approx$\SI{40}{\celsius}).
This saturation temperature increases to roughly \SI{75}{\celsius} if the A to C barrier is increased to 1.03~eV, and it decreases to \SI{0}{\celsius} in the case of the barrier being lowered to 0.83~eV.
While there is significant variation in the exact onset temperature, saturation is achieved in all three cases at $<$\SI{100}{\celsius}, a strong contrast to B$_2$H$_6$ and to a lesser extent PH$_{3}$.
The predictions in Fig.~\ref{fig:low-t-dosing} are compared to an experimental value (star) of the hole concentration ($p$) for a sample dosed with BCl$_{3}$ at room temperature and not annealed measured using Hall bar magnetotransport measurements.
%Given the reasonable agreement between the experimentally measured value and the originally calculated barrier of 0.93~eV, we will use the 0.93~eV value for the remainder of the simulations discussed here.

\begin{figure*}
\includegraphics[width=\textwidth]{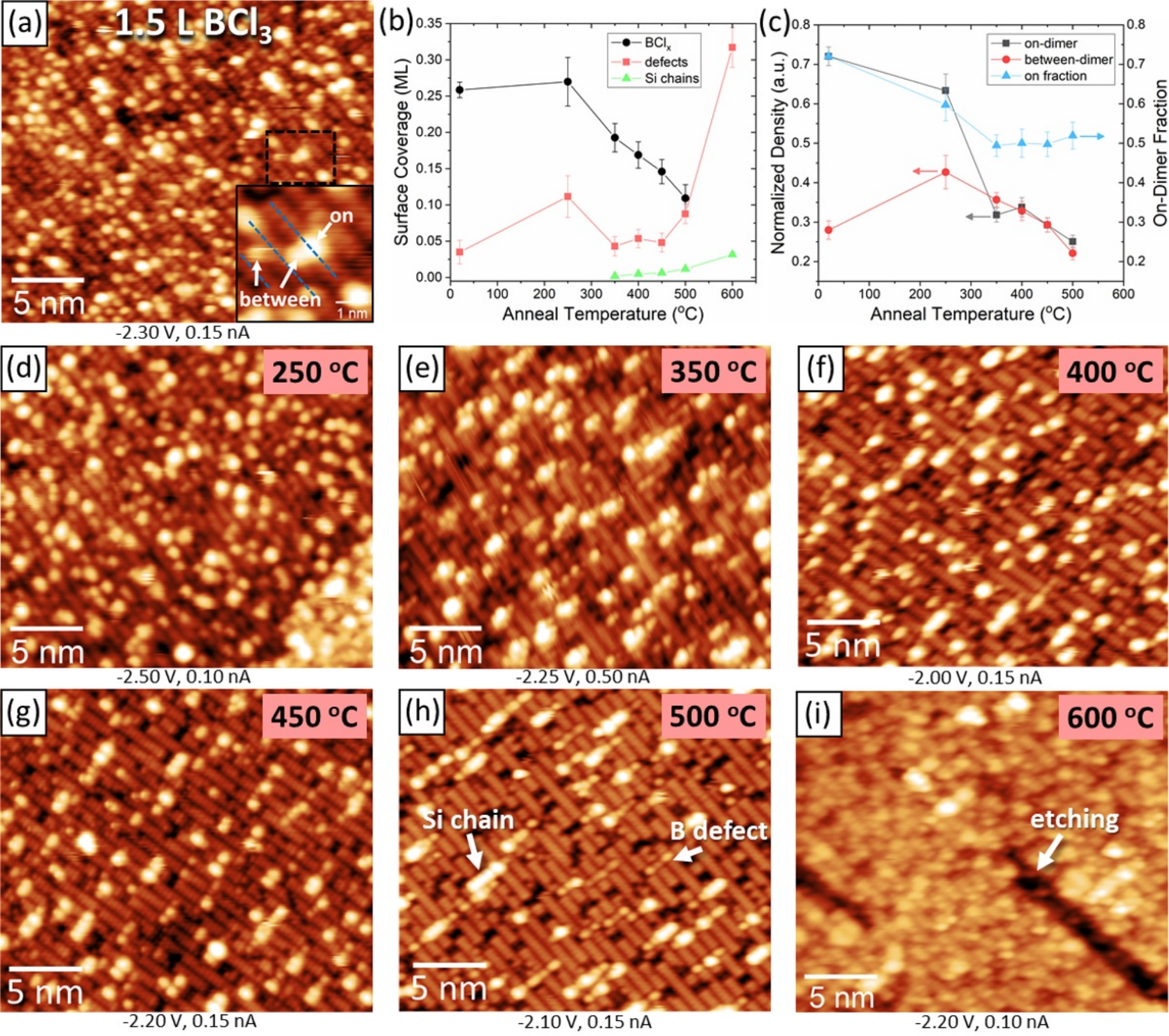}
\caption{Filled-state STM images of a Si(100) surface (a) dosed with 1.5~langmuir BCl$_{3}$ at room temperature and similar surfaces after annealing at (d) \SI{250}{\celsius}, (e) \SI{350}{\celsius}, (f) \SI{400}{\celsius}, (g) \SI{450}{\celsius}, (h) \SI{500}{\celsius}, and (i) \SI{600}{\celsius}. The inset of (a) magnifies the dashed box area and shows adsorbate features on and between dimer rows (dashed lines). At \SI{500}{\celsius}, Si chains and three-dimer-wide ``B defect'' features are observed (arrows), indicating B incorporation, and Cl etching is apparent at \SI{600}{\celsius}. (b) The surface coverage of BCl$_{x}$ adsorbates (circles), surface defects (squares), and Si chains (triangles) versus annealing temperature. Adsorbate coverage decreases with increasing temperature while defects and Si chains increase. (c) The normalized density of BCl$_{x}$ species (left axis) versus annealing temperature with on-dimer features (squares) converting to between dimers (circles). The on-dimer fraction (triangles, right axis) is also shown. Error bars are the standard error of feature counts.
\label{fig:STM}}
\end{figure*}

To verify the KMC predictions, STM was used to investigate the evolution of BCl$_{x}$ species adsorbed on the Si(100) surface as a function of annealing temperature. Figure~\ref{fig:STM} shows filled-state STM images of clean Si(100) surfaces after BCl$_{3}$ dosing and annealing in UHV. A surface dosed with 1.5~langmuir BCl$_{3}$ at room temperature is shown in (a) and is representative of all dosed samples before annealing. Numerous bright BCl$_{x}$ adsorbates are observed both on and between Si dimer rows, which are marked with dashed lines in the inset of (a). The DFT calculations suggest BCl$_{3}$ dissociatively adsorbs onto the surface at room temperature in a barrierless process, and the variety of larger and smaller adsorbate features observed in Fig.~\ref{fig:STM}(a) support this notion. The saturation dose used here results in a surface coverage of adsorbates of 0.26(1)~ML at room temperature. Figure~\ref{fig:STM}(d-i) show dosed surfaces after annealing at \SI{250}{\celsius}, \SI{350}{\celsius}, \SI{400}{\celsius}, \SI{450}{\celsius}, \SI{500}{\celsius}, and \SI{600}{\celsius}, respectively, revealing that as the annealing temperature is increased, the amount of bright adsorbates is visibly reduced.

Figure~\ref{fig:STM}(b) shows the surface coverage of BCl$_{x}$ adsorbates (circles), surface defects (squares), and Si chains (triangles) for each annealing temperature. At \SI{250}{\celsius}, the surface coverage of BCl$_{x}$ remains similar to that after dosing at room temperature (0.26(1)~ML). However, above \SI{250}{\celsius} the BCl$_{x}$ coverage drops linearly to just 0.11(2)~ML at \SI{500}{\celsius}. This reduction is likely due to a combination of factors including desorption of BCl$_{2}$ fragments, as predicted by the model for temperatures above \SI{250}{\celsius}, as well as incorporation of B into the surface. We note that the appearance of some surface features are potentially due to subsurface B atoms rather than BCl$_{x}$ adsorbates.\cite{Liu2008} At \SI{600}{\celsius}, the surface features became more ambiguous and were ignored in the coverage plot. Above \SI{350}{\celsius}, short Si chains appear along with an increase in dark dimer defects, as seen in Fig.~\ref{fig:STM}(h, i) and plotted in Fig.~\ref{fig:STM}(b). Some defects can be attributed to Cl atoms on the surface shed from BCl$_{3}$, but appear too similar to other defects to distinguish here. However, additional surface vacancy defects are likely caused by Cl on the surface which then etches Si\cite{Agrawal:2007} and is more pronounced at \SI{600}{\celsius} in Fig.~\ref{fig:STM}(i).

The appearance of Si chains can indicate ejected Si due to B incorporation. Here, short Si chains first appear at \SI{350}{\celsius}, and longer ones are evident at \SI{500}{\celsius} (Fig.~\ref{fig:STM}(h)) and \SI{600}{\celsius} (not shown) with a coverage of 0.032(2)~ML at \SI{600}{\celsius}. Unlike for P incorporation from PH$_3$ dosing, the Si chain coverage cannot give an accurate count of incorporated B due to the Cl etching and roughening process. B incorporation is also evidenced by the three-dimer-wide ``B defect'' features identified in Fig.~\ref{fig:STM}(h) likely related to subsurface B, which is known to form similar defects in highly B-doped Si(100) substrates.\cite{Liu2008, Zhang1996} Annealing at \SI{1000}{\celsius} leads to further B incorporation and diffusion into the surface causing the formation of the known ``bowtie'' appearance, affirming bulk-like B incorporation (see Supplementary Material).

To further compare the predicted incorporation path of the KMC model to experiment, STM images were analyzed to determine on-dimer and between-dimer feature statistics. From Fig.~\ref{fig:rxn_pathway_schematic}(a), BCl$_{2}$ fragments are expected to reside on one side of a dimer row (configuration A), while BCl is expected to protrude between dimer rows (configuration C). Simulated STM images of these two configurations indicate that both configuration A and C would appear as several distinct features (see Supplementary Material) with A consisting of a pair of bright features (BCl$_{2}$) on the dimer row and C consisting of several features (Cl) on the dimer row as well as a feature (BCl) centered between rows. As predicted by the model, 0.56 of the total B on the surface should reside in configuration C at room temperature, and so we estimate that $\approx$0.74 of the total features are expected to reside on dimer rows with $\approx$0.26 of the features being between rows. Figure~\ref{fig:STM}(c) plots the on-dimer (squares) and between-dimer (circles) density of features at each annealing temperature (left axis). The values are normalized to the initial total coverage to account for the decrease in adsorbate coverage with temperature. Also shown are the on-dimer features as a fraction of the total (triangles, right axis). At room temperature (no anneal), we find that 0.72(1) of the total features reside on dimer rows, in good agreement with the fraction estimated from the model (0.74), while 0.28(1) of features are between dimers. Thus we associate configuration A with several on-dimer features and C with both on-dimer and between-dimer features.

Above \SI{300}{\celsius}, the density of on-dimer features drops significantly to 0.32(2), similar to the between-dimer density. This drop can be attributed to a combination of BCl$_{2}$ desorption and full conversion of BCl$_{2}$ fragments in configuration A to BCl in C, predicted above \SI{300}{\celsius}. The on-dimer fraction remains constant at $\approx$0.5 at temperatures above \SI{350}{\celsius} as configuration C contributes some on-dimer features (Cl) as well. The increase in the between-dimer features from 0.28(1) of the initial total at room temperature to 0.43(4) at \SI{250}{\celsius} is also consistent with this interpretation.

While our proposed model has good agreement with experimental trends, there are still a number of factors that could be included for a more realistic picture of delta doping with BCl$_3$. 
The model does not account for site blocking by other adsorbates from the vacuum.
Additionally, the model only includes interactions along a Si dimer row, and not between rows.
While this serves as a good first approximation, interaction between dimer rows could result in decreased incorporation rates by enabling Cl atoms to migrate to adjacent rows but not BCl$_{x}$ fragments, thus limiting space on nearby rows for dissociation. Further, due to the increase in computational complexity, we have not considered any dissociation steps beyond the bridging BCl fragment. 
Nevertheless, the results in this work demonstrate that this simplified model works well as a first approximation of BCl$_3$ dissociation on the surface.
This helps confirm our dissociation pathway which is drastically simplified in comparison with PH$_3$.

\section{Conclusion}

In this work, we elucidate the dissociation pathway for BCl$_3$ using DFT, observing that BCl$_3$ adsorbs dissociatively onto the Si(100) surface in a barrierless process, forming a strong Si-Cl bond.
Based on this reaction pathway, we parameterize a KMC model and use it to predict incorporation rates as a function of dosing/annealing conditions.
We predict BCl$_3$ can incorporate to levels nearing $\approx$\SI{1.7e14}{cm^{-2}} with only room-temperature dosing and no further annealing step, corroborating this model with electrical transport measurements of BCl$_3$-doped $\delta$-layers.
STM is used to image BCl$_3$-dosed surfaces under these conditions, identifying surface structures and broadly verifying the predicted dissociation kinetics.
This work demonstrates a straightforward, thermodynamically downhill dissociation pathway for BCl$_3$ which presents a unique set of characteristics for creating $\delta$-doped layers in comparison with PH$_3$.
% We also identify a pathway for surface migration of BCl fragments at elevated temperatures, which can lead to a stable, electrically inactive dimer complex.
% Transport measurements of annealed $\delta$-layers also show a decrease in carrier concentrations at annealing temperatures $>$\SI{250}{\celsius}, however, SIMS measurements indicate that B dimerization does not play a significant role in the measured samples.
% Further, delaying the anneal step until after initial Si deposition results in the loss of
Our work creates a better understanding of BCl$_3$ characteristics for the creation of APAM-based bipolar devices, acceptor qubits, and superconducting regions within Si.

\section{Supplementary Material}

See the supplementary material/appendices for simulated STM images of BCl$_x$ adsorbates and an experimental STM image of BCl$_3$-dosed Si surface after incorporation at \SI{1000}{\celsius}.

\begin{acknowledgements}

The authors acknowledge S. Schmucker, J. Ivie, and J. Owen for useful discussions regarding modeling directions. The authors also thank J. Foglebach, D. Ketchum, P. Hannah, J. Moody, S. Brown and T. Olver for their assistance with BCl$_{3}$ handling and installation. Electrical measurements were supported by the Maryland Quantum Materials Center (facilities support).

This work was supported by the Laboratory Directed Research and Development program at Sandia National Laboratories under project 213017 (FAIR DEAL) and performed, in part, at the Center for Integrated Nanotechnologies, an Office of Science User Facility operated for the U.S. Department of Energy (DOE) Office of Science.
Sandia National Laboratories is a multi-mission laboratory managed and operated by National Technology and Engineering Solutions of Sandia, LLC, a wholly owned subsidiary of Honeywell International, Inc., for DOE’s National Nuclear Security Administration under contract DE-NA0003525.
This paper describes objective technical results and analysis.
Any subjective views or opinions that might be expressed in the paper do not necessarily represent the views of the U.S. Department of Energy or the United States Government.

\end{acknowledgements}

\section{Data Availability}
The data that support the findings of this study are available within the article and its supplementary material or from the corresponding author upon reasonable request. KMC code used in this study is available on github.\cite{kmccode}.

\input{BCl3_Annealing_Supplementary.tex}

\bibliography{extrabib}
\end{document}

%% file: BCl3_Annealing_Supplementary.tex
\appendix

\section{Simulated STM Images of BCl$_x$}

After equilibrium geometries were calculated using DFT, we generated an STM output with a sample bias of --2.2 V using QUANTUM ESPRESSO post processing software. We then use the CRITIC2 software package\cite{otero2014critic2} to generate the simulated constant current STM images with a threshold of 0.00005 atomic units of a 2$\times$2 supercell of the simulated surface.

\begin{figure*}[t]
\includegraphics[width=0.45\columnwidth]{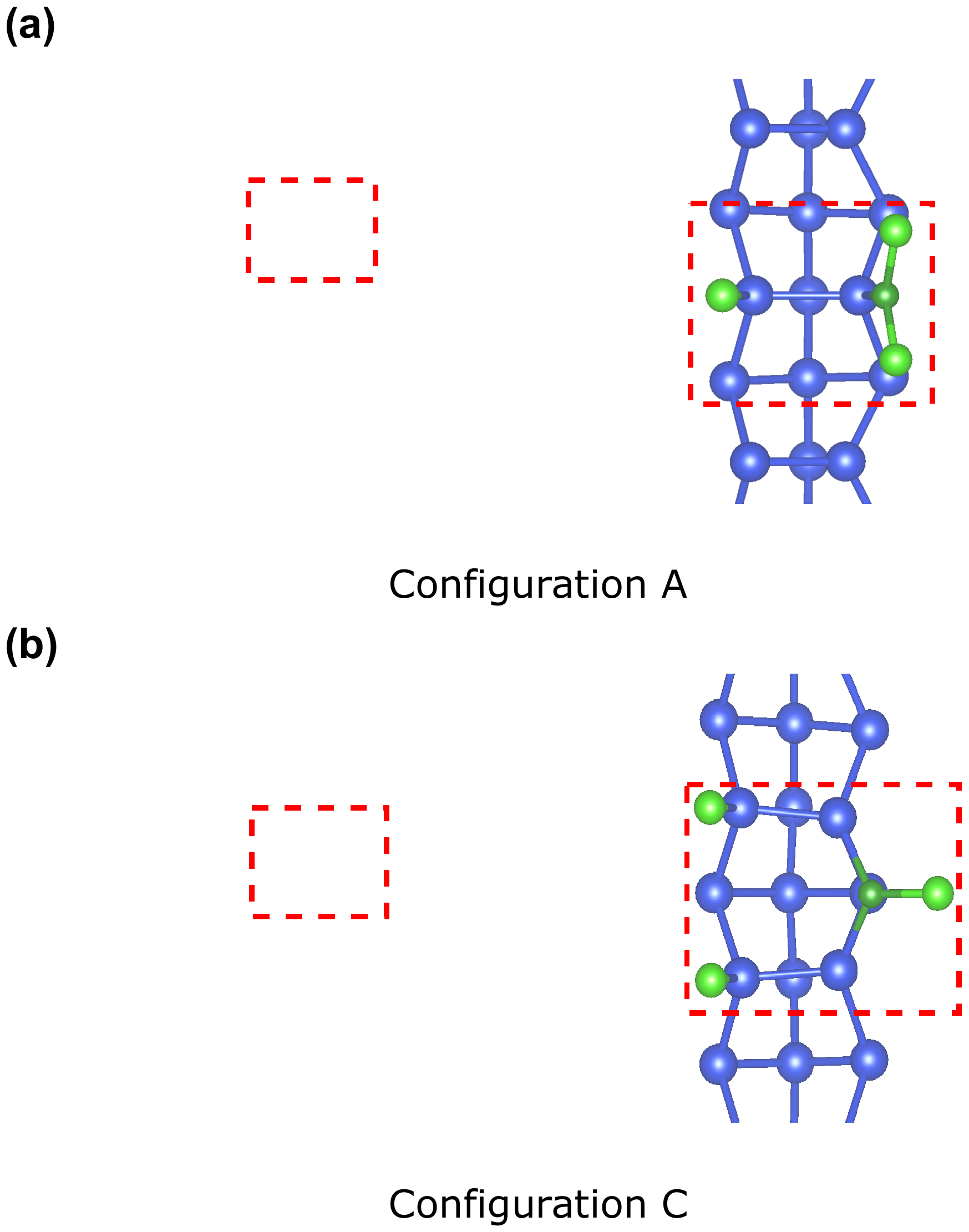}\hfill%
\caption{Simulated STM images of (a) configuration A and (b) configuration C, as labeled in Fig. 1 (main text). STM images are generated with a bias of --2.2 V. The dotted red box shows the correspondence between the atomic structure and the STM feature.}%
\label{fig:STM_sim}%
\end{figure*}

\section{STM of BCl$_3$-Dosed Si(100) Annealed at \SI{1000}{\celsius}}

\begin{figure*}[h!]
\includegraphics[width=0.35\columnwidth]{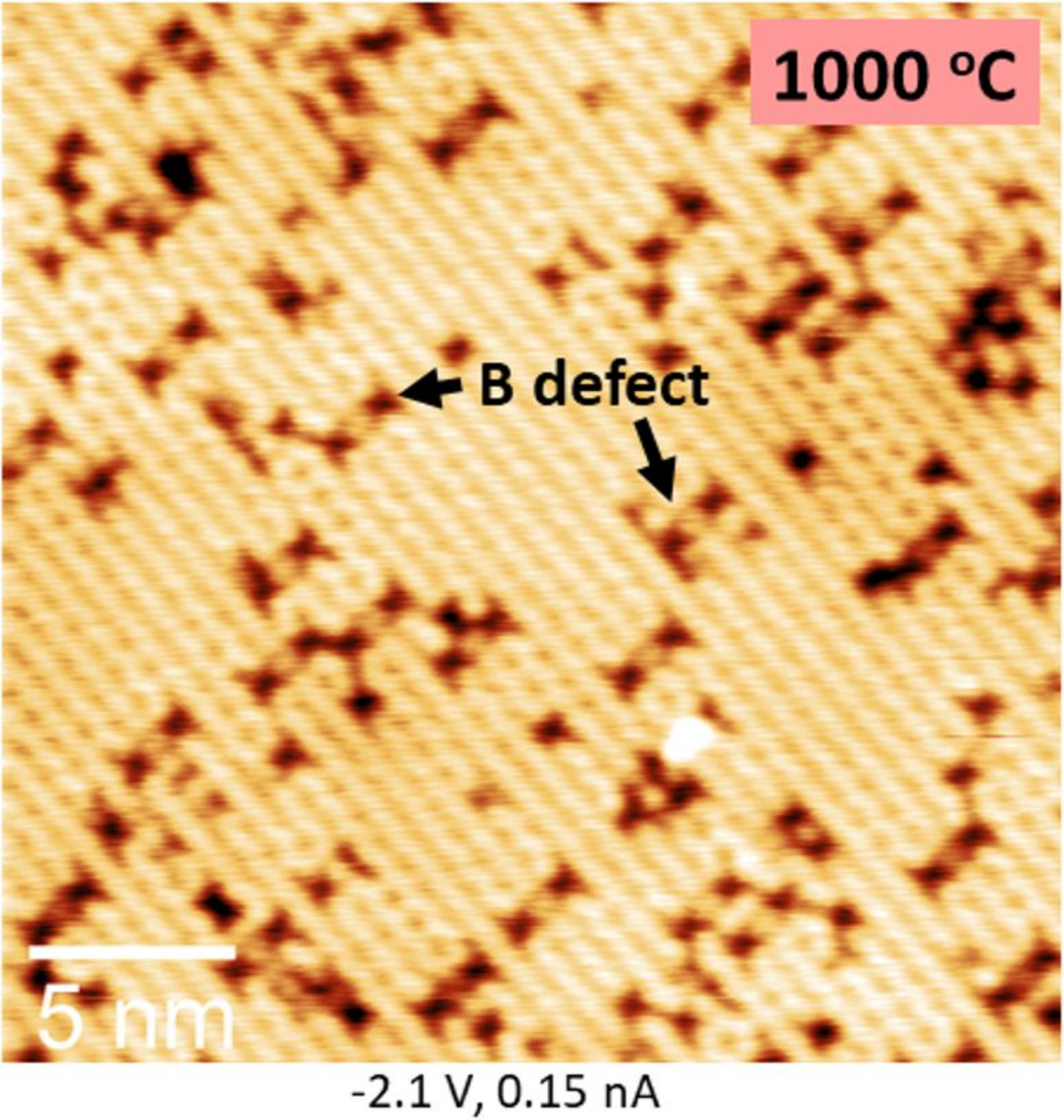}\hfill%
\caption{Filled-state STM images of a Si(100) surface dosed with 1.5 langmuir BCl$_{3}$ at room temperature and annealed at \SI{1000}{\celsius} for 1 min. Numerous three-dimer-wide ``bowtie'' defects typical of a highly B-doped substrate\cite{Liu2008} are visible on the surface indicating that B adsorbates have diffused several layers into the surface.}%
\label{fig:STM_1000C}%
\end{figure*}

\clearpage